\documentstyle[aps,prl,twocolumn]{revtex}
\input{epsfig}
\begin{document}
\draft
\title{Surface Screening Charge and Effective Charge}
\author{S. Clarke and J.E.Inglesfield}
\address{Department of Physics and Astronomy,University of Wales Cardiff,
P.O. Box 913, Cardiff CF2 3YB, United Kingdom}
\author{M. Nekovee}
\address{Department of Physics, The Blackett Laboratory, Imperial College, 
London SW7 2BZ, United Kingdom}
\author{P.K. de Boer}
\address{Department of Theoretical Physics, University of Nijmegen,
Toernooiveld, NL-6525 ED Nijmegen, The Netherlands}
\date{\today}
\maketitle

\begin{abstract}
The charge on an atom at a metallic surface in an electric field is defined as 
the field-derivative of the force on the atom, and this is consistent with
definitions of effective charge and screening charge. This charge can be found 
from the shift in the potential outside the surface when the atoms are moved.
This is used to study forces and screening on surface atoms of 
Ag(001) c$(2\times 2)$ -- Xe as a function of external field.
It is found that at low positive (outward) fields, the Xe with a negative
effective charge of $-0.093$ $|\mbox{e}|$ is pushed into the surface. 
At a field of 2.3 V \AA$^{-1}$ the charge changes sign, and for fields
greater than 4.1 V \AA$^{-1}$ the Xe experiences an outward force. Field
desorption and the Eigler switch are discussed in terms of these results.
\end{abstract}


The force on surface and adsorbate atoms in an applied electric field is important
for understanding field evaporation \cite{kreuzer}, 
field-induced reconstructions \cite{magnus}, and 
the movement of adsorbate atoms by STM tips \cite{eigler,pedro}. 
In the limit of low field, 
the force normal to the surface on atoms of type $i$ is given by:
\begin{equation}
F_{i}=q_{i}^{\ast}\mathcal{E},
\end{equation}
where $q_{i}^{\ast}$ is the effective charge on atom $i$ \cite{trull,mills,hamann,eac}. 
$\mathcal{E}$ is the external electric field above the surface, far away enough from
the atoms so that it is uniform -- this formula builds in all the local field effects
and screening at the surface. 
The effective charge is also responsible for the change
in work function $\phi$ when atoms of type $i$ are displaced\cite{hamann,eac}:
\begin{equation}
\frac{\partial\phi}{\partial z_{i}}=-\frac{4\pi q_{i}^{\ast}}{{\mathcal A}_{i}}
\end{equation}
-- here $z_{i}$ is the displacement of the atoms perpendicular to the surface, and
${\mathcal A}_{i}$ is the area per atom $i$. The surface effective charge is already 
familiar to us from the theory of the interaction of probes such as EELS and
IR with surface vibrations \cite{mills}, and it is the same as the Born
effective charge in polar solids in which there is much current interest
\cite{vdb,resta}.
In this letter we shall generalize
these ideas to calculate the force on an atom 
at a metallic surface in an electric field 
of arbitrary strength. This will allow us to determine not only the effective
charge, but also to assign the screening charge to individual atoms.

We use an extension of the classical argument for finding the force
on the plate of a parallel plate capacitor -- one plate consists of the
surface under consideration, and the other is an arbitrary electrode.
The potential of the electrode is initially $V$ relative to our surface,
which is maintained at constant potential, and the two are disconnected.
On taking charge $dq$ from the
electrode to the surface the energy change of the system is $dU$, with:
\begin{equation}
V=-\frac{\partial U}{\partial q}.
\end{equation}
If we now move the atoms of type $i$ by $dz_i$, the potential across the capacitor
changes (figure 1), in just the same way that the work function changes in the 
zero-field case. From (3) we have:
\begin{equation}
\frac{\partial V}{\partial z_{i}}=-\frac{\partial^{2}U}{\partial q\partial z_{i}}.
\end{equation}
But $-\partial U/\partial z_{i}$ is the total force on the atoms of type $i$, 
so (4) becomes:
\begin{equation}
\frac{\partial V}{\partial z_{i}}=N_{i}\frac{\partial F_{i}}{\partial q},
\end{equation}
where $F_{i}$ is the force on each atom of type $i$, and there are $N_{i}$
such atoms on the surface. Replacing $dq$ by the change in the electric field
$d\mathcal{E}$ between the plates we obtain a relationship between
$\partial F_{i}/\partial\mathcal{E}$ and $\partial V/\partial z_{i}$:
\begin{equation}
{\mathcal Q}_{i}\equiv\frac{\partial F_{i}}{\partial\mathcal{E}}=
\frac{{\mathcal A}_{i}}{4\pi}\frac{\partial V}{\partial z_{i}}.
\end{equation}
We \emph{define}
$\partial F_{i}/\partial\mathcal{E}$ as the charge ${\mathcal Q}_{i}$ on atoms $i$, 
and we can then determine it from the right-hand side of (6). (The difference in
sign between (6) and (2) arises because $V$ is the electrostatic potential,
whereas $\phi$ is the electron potential energy.) The right-hand side of this
expression is easy to evaluate in an electronic structure calculation in the
presence of an applied field -- we simply have to move the atoms of type $i$ and
see how much the potential in the vacuum shifts.

${\mathcal Q}_{i}$ is also related to the shift of the centre of gravity of the
screening charge with atomic displacement. The centre of gravity $z_{0}$ is the
electrostatic origin of the electric field \cite{lang,aers}, 
which means that:
\begin{equation}
\frac{\partial V}{\partial z_{i}}={\mathcal E}\frac{\partial z_{0}}{\partial z_{i}}.
\end{equation}
But $\mathcal{E}$ is related to the total screening charge $\mathcal{Q}$ per surface
unit cell area $\mathcal{A}$ by:
\begin{equation}
{\mathcal E}=\frac{4\pi\mathcal{Q}}{{\mathcal A}},
\end{equation}
so from (6):
\begin{equation}
{\mathcal Q}_{i}=\frac{{\mathcal A}_{i}{\mathcal Q}}{{\mathcal A}}\cdot
\frac{\partial z_{0}}{\partial z_{i}}.
\end{equation}
Now ${\mathcal A}_{i}/{\mathcal A}=1/{\mathcal N}_{i}$, 
where ${\mathcal N}_{i}$ is the number of atoms of
type $i$ per unit cell, so the charge associated with these atoms is given by:
\begin{equation}
{\mathcal Q}_{i}=\frac{{\mathcal Q}}{{\mathcal N}_{i}}\cdot
\frac{\partial z_{0}}{\partial z_{i}}.
\end{equation}
This is an obvious way to divide the charge on the surface between the different
atoms, and we conclude that (6) is a natural and unambiguous definition of
${\mathcal Q}_{i}$.

By moving all the surface atoms through $dz$ we simply shift the total screening 
charge, hence:
\begin{equation}
\sum_{i}\frac{\partial z_{0}}{\partial z_{i}}=1.
\end{equation}
So we have the sum-rule:
\begin{equation}
\sum_{i} {\mathcal N}_{i}{\mathcal Q}_{i}={\mathcal Q},\label{sumrule}
\end{equation}
and the charge on all the atoms adds up to the total screening charge. If we
separate ${\mathcal Q}_{i}$ into the zero-field effective charge $q_{i}^{\ast}$ plus
a term linear in the external field:
\begin{equation}
{\mathcal Q}_{i}=q_{i}^{\ast}+\alpha_{i} {\mathcal E},
\end{equation}
we see from (6) that the force on atoms of type $i$ in the field is given by:
\begin{equation}
F_{i}=q_{i}^{\ast}{\mathcal E}+\frac{1}{2}\alpha_{i} {\mathcal E}^{2}
\end{equation}
-- the factor of $1/2$ is familiar to us from the force on a capacitor plate.
In these expressions for the force, we have implicitly assumed that
the atoms have not been allowed to relax from their zero-field positions; 
what actually happens is that the atoms move in the field so that this
electrostatic force is balanced by the interatomic forces, in this way 
distributing the force as a stress throughout the system.

To show how these ideas can give useful information about bonding and screening,
we consider Ag(001) c$(2\times 2)$ -- Xe, with the Xe atoms adsorbed
in atop sites. We have performed self-consistent electronic structure calculations
for this system, using the embedding method 
\cite{segf} to include the semi-infinite substrate
and treating the top two layers of atoms explicitly. We calculate the charges
${\mathcal Q}_{i}$ for a range of applied electric fields from (6), by 
moving the atoms through small displacements (typically 
0.02 a.u.) and calculating the change in electrostatic potential
outside the surface. Our results are shown in table 1, and first we note that for
all the fields the sum of charges ${\mathcal Q}_{i}$ is very close to the total
screening charge ${\mathcal Q}$ calculated from (8) -- the sum-rule (\ref{sumrule}) is
well satisfied. The slight discrepancies are due to the fact that the screening
is confined to the top two layers of atoms in our calculation.

In the limit of zero field, we find that Xe has a small
negative effective charge, $q^{\ast}_{\mbox{\footnotesize Xe}}=-0.093$ $|\mbox{e}|$, 
which is largely counterbalanced by a positive effective
charge on the underlying Ag atoms. The sign of the effective charge agrees with
field emission experiments, which show that a positive field (in our convention, 
one which corresponds to positive screening charge at the surface) pushes inert 
gas atoms towards the surface \cite{ernst}.

The negative effective charge on the Xe can be understand in terms of the dipole
moment of adsorbed rare gas atoms and its variation with distance \cite{denis}. 
Adsorbed Xe has
a dipole moment pointing out of the surface (positive charge outside, negative
charge inside), as is shown by the decrease in work function compared with the
clean surface \cite{merry}. 
The physical origin of the positive dipole moment lies in 
the fact that the vacuum side is less attractive to electrons on the Xe atoms
than the metal side, where the exchange-correlation 
potential is more attractive \cite{lang2}. Now a Van der
Waals treatment gives a dipole moment varying as $(z_{\mbox{\footnotesize Xe}}
-z_{VdW})^{-4}$ \cite{zaremba}, where $z_{VdW}$ is the position of the Van der Waals plane, 
so as $z_{\mbox{\footnotesize Xe}}$ increases and the
Xe atoms are moved away from the surface, the dipole moment decreases. The
work function increases, and this corresponds to a negative effective charge on
the Xe atoms. This approach gives:
\begin{equation}
q^{\ast}_{\mbox{\footnotesize Xe}}\approx\frac{{\mathcal A}_{i}}{\pi}\frac{\Delta\phi}
{(z_{\mbox{\footnotesize Xe}}-z_{VdW})},
\end{equation}
where $\Delta\phi$ is the change in work-function due to Xe adsorption.
Taking $(z_{\mbox{\footnotesize Xe}}-z_{VdW})=4.2$ a.u., $\Delta\phi=-0.46$ eV, 
and ${\mathcal A}_{i}=60.7$ a.u., we then obtain an effective charge of 
$-0.08$ $|\mbox{e}|$, in excellent agreement with our first-principles
calculation. However, this is not the whole story, as a tight-binding
calculation of Xe on Al shows some chemisorption, with real charge transfer of
about $-0.1$ $|\mbox{e}|$ to the Xe \cite{perez}.
Whatever the origin of the effective charge of 
$-0.093$ $|\mbox{e}|$ on the Xe, it is this charge which determines the force in the 
low field limit.

As the electric field is turned on, the charge on the Xe atoms becomes slightly
less negative, whereas there is a much bigger increase in positive charge on the
Ag atoms, particularly Ag(1) -- the Ag atom directly underneath the Xe. 
These results mean that
the screening charge is mostly situated on the surface Ag atoms, in agreement
with our expectations. What is surprising is that Ag(1) is responding twice as
much to the electric field as Ag(2), and there are several possible explanations. 
First, the Xe lowers the Ag work-function, and if this is considered as a local 
effect we would expect that the valence wavefunctions on the Ag atoms under the
Xe should penetrate deeper into the vacuum
\cite{eigler2}
and hence respond more strongly to external fields. Secondly, it may be a local 
field effect with enhancement of the field under the Xe atoms. 

The results for the charge on the Xe shown in table 1, 
extended to larger electric fields up to
${\mathcal E}=0.06$ a.u., can be fitted by the formula:
\begin{equation}
{\mathcal Q}_{\mbox{\footnotesize Xe}}=-0.097+1.165{\mathcal E}+23.147{\mathcal E}^{2}
\qquad \mbox{(a.u.)}.\label{charge}
\end{equation}
At a field of 0.044 a.u. (2.3 V \AA$^{-1}$), the field-induced charge cancels the
effective charge, and the charge on the Xe atoms changes sign.
The force on the Xe is given by integrating (\ref{charge}) with respect to 
${\mathcal E}$ (6):
\begin{equation}
F_{\mbox{\footnotesize Xe}}=-0.097{\mathcal E}+0.583{\mathcal E}^{2}+7.715{\mathcal E}^{3}
\qquad \mbox{(a.u.)}.\label{force}
\end{equation}
In a positive electric field, the Xe atoms feels a force towards the surface at low fields. 
The force towards the surface initially increases linearly with increasing field, 
but then decreases, going through zero at ${\mathcal E}=0.08$ a.u. 
(4.1 V \AA$^{-1}$) and directed away from the surface for larger fields. 
We emphasise that this is the total force on the Xe atoms maintained at the 
original  position, and it takes into account all the field-induced
bonding and charging \cite{wang}, and local field enhancement effects. 
The change in sign of the force at large fields agrees with cluster
calculations of adsorbed rare gas atoms by Nath et al. \cite{nath}. 
The rapid variation of $F_{\mbox{\footnotesize Xe}}$ at large $\mathcal{E}$ 
means that a field somewhat greater than 4.1 V \AA$^{-1}$ will overcome
the Van der Waals and other bonding forces and remove
Xe atoms from the Ag(001) surface. In fact this certainly overestimates
the field needed for field desorption  -- it does not take account 
of the non-adiabatic transition to ionized Xe.

The fields necessary for field desorption 
are much greater than those involved in the manipulation of surface
Xe atoms by the STM. In the Eigler switch \cite{eigler3}, 
in which a positively biased tip
pulls a Xe atom from the surface and the reverse bias switches it back, typical
parameters are a tip bias of $+0.8$ V with the tip about 4 \AA\ above the
surface. This corresponds to an electric field $\mathcal{E}$ of $-0.004$ a.u. 
(in our sign convention),  so from (\ref{charge}) we see that ${\mathcal Q}_
{\mbox{\footnotesize Xe}}$ is essentially the effective charge. This field
pulls the Xe towards the tip with a force of $4\times 10^{-4}$ a.u., 
20 meV \AA$^{-1}$, much smaller than the force needed to pull a Xe atom directly
off the surface (from potential energy curves this is 100-150 meV \AA$^{-1}$ 
\cite{bouju,mingo}). As shown by de Andres et al.\cite{pedro} and  Walkup et al.
\cite{denis}, the importance of the force on the effective charge is that it 
produces a bias-dependent shift of the potential well of the Xe adsorbed on the 
tip relative to the well for adsorption on the surface, facilitating the Xe atom 
jumps. 

We thank H.J. Kreuzer for helpful correspondence, and R.G Forbes for useful
discussions.

\begin{figure}
\epsfxsize=10cm
\centerline{{\epsffile{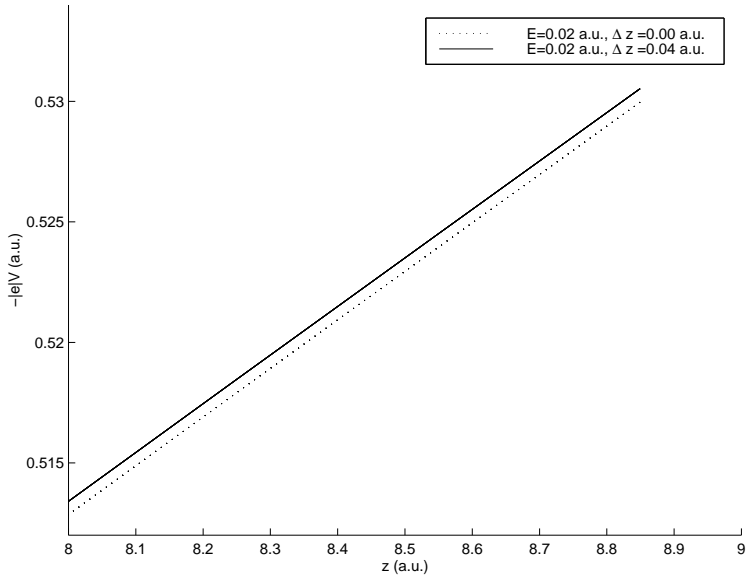}}}
\caption{Shift in electron potential energy outside the surface in an 
electric field of +0.02 a.u., on displacing Xe atoms by 0.04 a.u. into the vacuum.
The undisplaced Xe atoms are at +2.754 a.u.}
\end{figure}

\begin{table}
\begin{center}
\begin{tabular}{lcccc}
&${\mathcal E}=0$&$+0.005$&$+0.01$&$+0.02$\\ \tableline
Xe&$-0.093$&$-0.089$&$-0.084$&$-0.069$\\
Ag(1)&$+0.086$&$+0.099$&$+0.115$&$+0.136$\\
Ag(2)&$+0.002$&$+0.008$&$+0.016$&$+0.031$\\ \tableline
${\mathcal Q}$&0&+0.024&+0.048&+0.097
\end{tabular}
\end{center}
\caption{Charge on atoms at Ag(001) c$(2\times 2)$ -- Xe, for different electric 
fields. The Xe atoms are atop Ag(1). The last row shows the total screening charge 
per surface unit cell calculated from (8). Charges 
are in units of $|\mbox{e}|$, and fields in atomic units; a positive field is
directed out of the surface.}
\end{table}
\end{document}